\documentclass[pra, twocolumn, amsmath, amssymb, notitlepage, longbibliography, showpacs, superscriptaddress]{revtex4-1}

\usepackage{graphicx}
\usepackage{dcolumn}
\usepackage{bm}

\usepackage{textcomp} 
\usepackage{dsfont}
\usepackage[english]{babel}
\usepackage{lipsum}
\usepackage[usenames, dvipsnames]{color}
\usepackage{rotating}
\usepackage[breaklinks=true]{hyperref}
\hypersetup{
	colorlinks   = true, 
	urlcolor     = blue, 
	linkcolor    = blue, 
	citecolor   =  red 
}

\graphicspath{ {./fig/} }

\newcommand{\bra}[1]{\ensuremath{\left\langle #1\r|}}
\newcommand{\ket}[1]{\ensuremath{\left|#1\r\rangle}}

\newcommand{\mean}[1]{\ensuremath{\left\langle #1\r\rangle}}

\newcommand{\cc}{^{\ast}}							
\newcommand{\hc}{^{\dagger}}							
\newcommand{\HC}{\textrm{h.c.}}
\newcommand{\ee}{\mathrm{e}}						
\newcommand{\ii}{\mathrm{i}}			             			

\newcommand{\comm}[2]{\left[ #1, #2 \right]} 				
\newcommand{\nn}{\nonumber}							
\newcommand{\abss}[1]{\ensuremath{ \left| #1 \right|^{2} }}	
\newcommand{\diss}[1]{\mathcal{D}[ #1 ]}					
\renewcommand{\l}[0]{\left}
\renewcommand{\r}[0]{\right}

\newcommand{\Tr}{\text{Tr}}

\newcommand{\overbar}[1]{\mkern 1.5mu\overline{\mkern-1.5mu#1\mkern-1.5mu}\mkern 1.5mu}

\usepackage{epstopdf} 
\DeclareGraphicsExtensions{.pdf,.png} 

\begin{document}

\title{Nonreciprocity realized with quantum nonlinearity}

\author{Andr\'es {Rosario Hamann}}
\email{arosario@uq.edu.au}
\affiliation{ARC Centre of Excellence for Engineered Quantum Systems, School of Mathematics and Physics, The University of Queensland, Saint Lucia, Queensland 4072, Australia}

\author{Clemens M\"uller}
\affiliation{ARC Centre of Excellence for Engineered Quantum Systems, School of Mathematics and Physics, The University of Queensland, Saint Lucia, Queensland 4072, Australia}
\affiliation{Institute for Theoretical Physics, ETH Z\"urich, 8093 Z\"urich, Switzerland}

\author{Markus Jerger}
\affiliation{ARC Centre of Excellence for Engineered Quantum Systems, School of Mathematics and Physics, The University of Queensland, Saint Lucia, Queensland 4072, Australia}

\author{Maximilian Zanner}
\affiliation{Physikalisches Institut, Karlsruhe Institute of Technology (KIT), 76131 Karlsruhe, Germany}

\author{Joshua Combes}
\affiliation{ARC Centre of Excellence for Engineered Quantum Systems, School of Mathematics and Physics, The University of Queensland, Saint Lucia, Queensland 4072, Australia}

\author{Mikhail Pletyukhov}
\affiliation{Institute for Theory of Statistical Physics, RWTH Aachen University, 52056	Aachen, Germany}

\author{Martin Weides}
\affiliation{Physikalisches Institut, Karlsruhe Institute of Technology (KIT), 76131 Karlsruhe, Germany}
\affiliation{School of Engineering, Electronics \& Nanoscale Engineering Division, University of Glasgow, Glasgow G12 8QQ, UK }

\author{Thomas M. Stace}
\affiliation{ARC Centre of Excellence for Engineered Quantum Systems, School of Mathematics and Physics, The University of Queensland, Saint Lucia, Queensland 4072, Australia}

\author{Arkady Fedorov}
\email{a.fedorov@uq.edu.au}
\affiliation{ARC Centre of Excellence for Engineered Quantum Systems, School of Mathematics and Physics, The University of Queensland, Saint Lucia, Queensland 4072, Australia}

\date{\today}

\begin{abstract}
Nonreciprocal devices are a key element for signal routing and noise isolation.  Rapid development of quantum technologies has
boosted the demand for a new generation of miniaturized and low-loss nonreciprocal components. Here we use a  pair of tunable superconducting 
artificial atoms in a 1D waveguide  to experimentally realize a minimal passive nonreciprocal device. 
Taking advantage of the quantum nonlinear behavior of artificial atoms, we 
achieve nonreciprocal transmission through 
the waveguide in a wide range of powers. 
Our results are consistent with theoretical modeling showing that nonreciprocity is associated with
the population  of the two-qubit nonlocal entangled quasi-dark state, 
which responds asymmetrically to incident fields from opposing directions. 
Our experiment highlights the role of quantum correlations in
enabling nonreciprocal behavior and opens a path to building 
passive quantum nonreciprocal devices without magnetic fields. 
\end{abstract}

\maketitle

Microwave nonreciprocal devices based on ferromagnetic compounds increase 
signal processing capabilities, but they are bulky and inherently lossy~\cite{Pozar1998}. 
Different approaches to achieve nonreciprocity on a chip are being actively pursued to enable 
circuits of greater complexity and advanced functionality. A common path to achieve  
nonreciprocity consists in breaking time-reversal symmetry, either by utilizing  
novel materials~\cite{Viola2014,Mahoney2017,Muller2018}
or by exploiting sophisticated time control 
schemes~\cite{Estep2014,Kerckhoff2015,Barzanjeh2017, Bernier2017, Fang2017,Chapman2017}. 
Here we follow another path and use a pair of tunable superconducting artificial atoms 
in a 1D waveguide in order to realize the simplest possible nonreciprocal device 
without breaking time-reversal symmetry. 
In contrast to isolators based on nonlinear bulk media 
response~\cite{Fan2012,Yi2015}, nonlinear resonances~\cite{Sounas2018}, or nonlinearity enhanced by active breaking of the parity-time symmetry~\cite{Peng2014},
our system exploits the quantum nonlinear behavior of a minimal system comprised of two 
two-level artificial atoms ~\cite{Fratini2014,Dai2015,Fratini2016,Muller2017a}. 
This quantum nonlinearity, combined with an asymmetric atomic detuning
that breaks the structural symmetry of the system, leads to population trapping 
of an entangled state and, ultimately, to 15~dB isolation 
in a wide range of powers controllable by the experimental settings. 
Our experiment provides insights into the role of quantum correlations in generating 
nonreciprocity and open a new path towards the realization of nonreciprocal quantum 
devices on a chip.


Schemes for building nonreciprocal devices based on nonlinearity of quantum emitters were first proposed in Ref.~\cite{Roy2010,Roy2013}.
A more specific implementation of a {\it quantum diode} built of two atoms in 1D open space 
was later proposed in Ref.~\cite{Fratini2014} and has attracted 
significant theoretical attention since ~\cite{Dai2015,Fratini2016,Mascarenhas2016,Fang2017, Muller2017a}. 
The quantum theory of the diode was first presented in 
Refs.~\cite{Dai2015, Fratini2016,Fang2017} and later work revealed 
the detailed mechanism of nonreciprocity, determining analytical bounds for the device efficiency and
identifying entanglement between the atoms and the electromagnetic field
as a crucial element in the nonreciprocal behavior of the system~\cite{Muller2017a}. 
In this work, we present experimental results on the realization of the 
quantum diode and provide compelling evidence of the connection of its 
nonreciprocity with the population of the entangled quasi-dark state.

\begin{figure}[t]
  \includegraphics[width=\columnwidth]{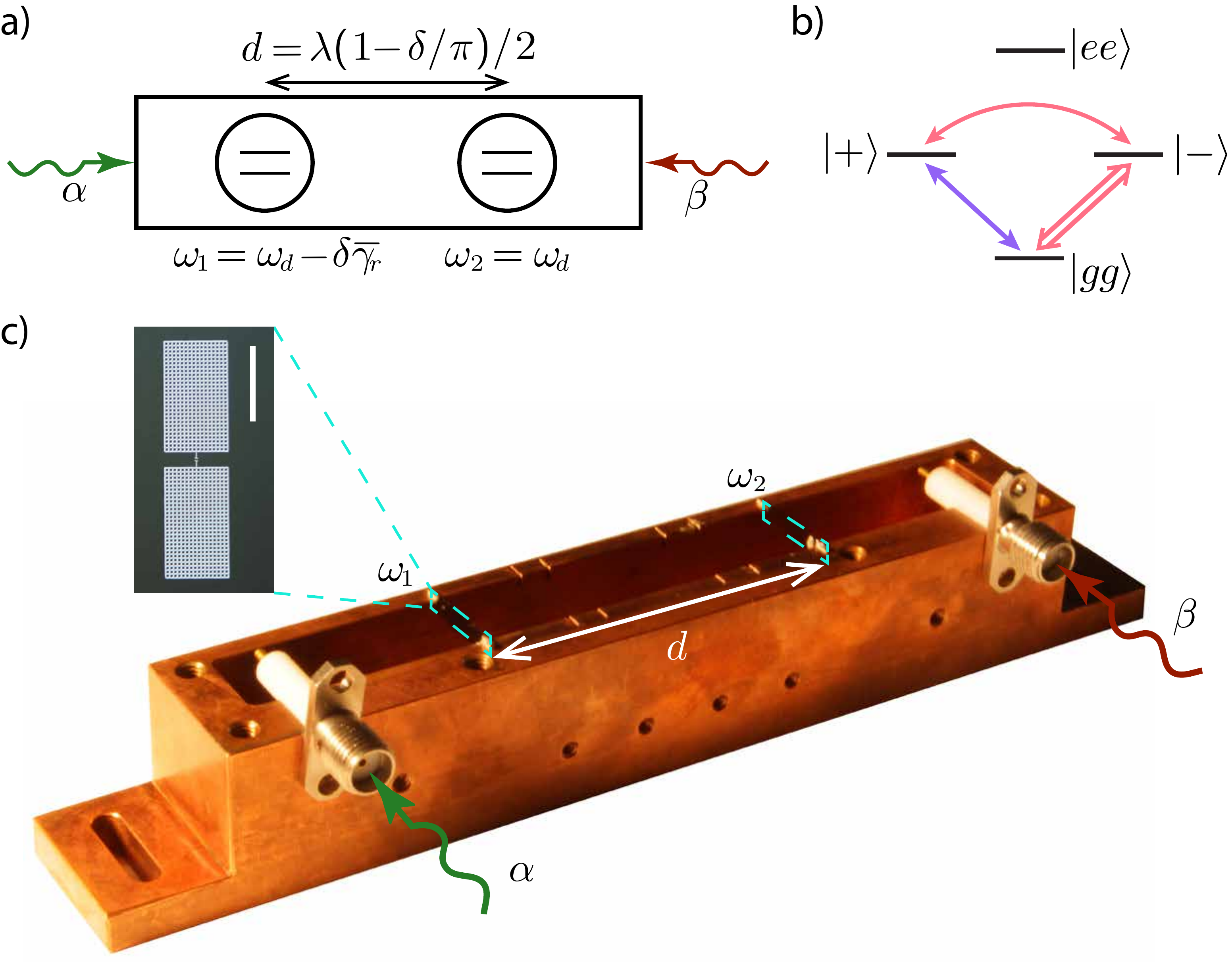}
    \caption{a) Schematic of the quantum diode: two qubits embedded 
             in a 1D waveguide, tuned to the optimal conditions for nonreciprocal behavior 
             ($\omega_1=\omega_d-\delta\overbar{\gamma_r}$, $\omega_2=\omega_d$).  
             An incoming field from the forward direction ($\alpha$ drive)
             at frequency $\omega_d$ is partially transmitted through 
             the system, whereas a field incoming from the reverse direction ($\beta$ drive) 
             is fully reflected.
             b) Energy level diagram of the system. The quasi-dark state 
             $\ket{+}$ can be populated by the driving field either directly from the 
             ground state $\ket{gg}$ (purple path) or indirectly through the bright state 
             $\ket{-}$ (pink path). These two channels interfere either constructively 
             or destructively depending on the driving direction. If interfering constructively,
             part of the population gets trapped in the quasi-dark state $\ket{+}$,
             which in turn gives rise to the nonreciprocal behavior of the system.
             c) Open 1D waveguide with embedded 3D transmons (dashed mint green boxes). 
             Inset: Optical micrograph of one of the two identical 3D transmons.
             The scale bar corresponds to 500~\textmu m. }
	\label{FigSchematic}
\end{figure}

More specifically, we use two transmon-type superconducting qubits 
inserted in a rectangular copper waveguide (see Fig.~\ref{FigSchematic}c). 
The qubits are spatially separated by $d=22.5~\,\textrm{mm}$
and are oriented to maximize coupling 
to the $\textrm{TE}_{10}$ mode, which has a lower cutoff at $f_{c,\,10}=6.55\textrm{ GHz}$.
Two microwave connectors are positioned near each end of the waveguide, providing 
an interface between the microwave field inside the waveguide and the external circuitry. 
This ensures that the qubits are coupled to the continuum of the electromagnetic 
modes, thus emulating an effective 1D open space. Further technical 
details into the transmons and 1D waveguide design and characterization
can be found in Ref.~\cite{Zanner3001}.

The qubits were patterned by standard electron-beam lithographic 
techniques on high resistivity Si, followed by two angle shadow evaporation of Al.
The design of the circuit consists of two planar capacitor plates connected 
via a line interrupted by a SQUID, playing the role of a tunable Josephson 
junction (see inset Fig.~\ref{FigSchematic}c). 
The transition frequencies of the qubits are then controlled 
via two current-biased superconducting coils. 

As a result of the interaction with the waveguide modes, the excited state of a qubit $\ket{e}$
spontaneously relaxes to its ground state $\ket{g}$ at the radiative decay rate 
$\gamma_{r}/2\pi$.
This interaction leads to an almost full reflection of incident resonant 
microwaves by the qubit at low powers~\cite{Astafiev2010}, a phenomenon  we use 
to determine the frequency of our qubits, their radiative decay rates and their 
decoherence due to other noise channels (refer to the Supplementary Material
for more details). The rates $\gamma_{r}/2\pi$  were found to 
depend on the transition frequencies of the qubits $f_{ge}$, and varied between 60~MHz and 
85~MHz for $f_{ge}$ between 8.5~GHz and 9.0~GHz, respectively. 
The transmittance at resonance with the qubit was extinguished to less than $0.4\%$ 
at low powers of incident radiation, providing an upper bound on the qubits' decoherence rate. 
This is characterized by the non-radiative decay $\gamma_{nr}$  and 
dephasing $\gamma_\phi$ rates, which we measured to fall below $0.5\%\,\gamma_r$
for both qubits (detailed values can be found in the Supplementary Material). 
In order to determine the dependence of $f_{ge}$ on the external magnetic 
field supplied by the coils, we performed transmission measurements while varying the 
magnetic field produced by each coil. We found the maximum frequencies of the qubits 
to be $9.9$~GHz and $11.0$~GHz.


When the two-atom system is driven by an external microwave field, 
their interaction depends strongly on the distance between qubits. 
Specifically, the interatomic distance $d$ determines the phase $\phi$ 
acquired by the drive when traveling from one qubit to the other, $\phi=\omega_d d/v_p$,
where $v_p$ is the phase velocity in the waveguide and
$\omega_d$ is the frequency of the drive. 
We tune $\phi$ \emph{in situ} by setting the frequency $\omega_d$ of the incoming drive.


This interaction between the two qubits with the continuum of 
the electromagnetic modes in the waveguide gives rise to
a field mediated exchange coupling between the qubits described via the term~\cite{Lalumiere2013}  
$H_C = \tfrac{1}{2}\,\overbar{\gamma_r} \sin\phi\left(\sigma_{-}^{(1)}\sigma_{+}^{(2)}+\HC\right)$, 
where $\sigma_{-}=\ket{g}\bra{e}$ and $\overbar{\gamma_r}\equiv\sqrt{\gamma_{r,1} \gamma_{r,2}}$
(see Supplementary Material).
At the phase matching condition $\phi=\pi$
(which, in the case of our system occurs when $\omega_d = \omega_\pi$,
with $\omega_\pi \equiv 8.975\textrm{ GHz}$),
the exchange coupling between the qubits vanishes, so that the symmetric and antisymmetric states 
$\ket{\pm} = \left(\ket{ge} \pm \ket{ge}\right)/\sqrt{2}$ are perfectly degenerate.
Furthermore, the antisymmetric state $\ket{-}$ is bright,
with a decay rate $\Gamma_-=2\overbar{\gamma_r}$, 
whereas the symmetric state $\ket{+}$ is dark, and hence fully decoupled from 
the interaction with the waveguide modes: $\Gamma_+=0$~\cite{Lalumiere2013}.

If the qubits are slightly detuned from the frequency $\omega_\pi$, 
a resonant field at $\omega_d$ acquires a phase $\phi=(\omega_d/\omega_\pi)\pi\equiv\pi-\delta$, where  the 
small parameter $\delta \ll 1$ characterizes the detuning from the phase matching condition. 
In this case, the exchange interaction between qubits does not vanish and 
lifts in turn the degeneracy between the $\ket{\pm}$  states: 
$H_C=(J/2)(\sigma_+^{(1)}\sigma_-^{(2)}+\HC)$,
with $J=\overbar{\gamma_r} \sin\phi\simeq \overbar{\gamma_r} \delta$. 
To leading order in $\delta$, the dark state $|+\rangle$ becomes quasi-dark 
with a decay rate $\Gamma_+=\delta^2 \overbar{\gamma_r}$, while 
the bright state decay rate remains unchanged: $\Gamma_-=2\overbar{\gamma_r}$,~\cite{Redchenko2014,Muller2017a}.

To break the inversion symmetry of our device and achieve nonreciprocal behavior, we set qubit 2
to be resonant with the incoming field, $\omega_2=\omega_d$,
whereas qubit 1 is set at $\omega_1=\omega_d-\delta\overbar{\gamma_r}$, 
to compensate for the phase asymmetry introduced by the detuned $\omega_d$ (see Fig.~\ref{FigSchematic}a).
This configuration opens an additional path of accessing
the quasi-dark state $\ket{+}$, which can now be populated 
either directly by the incoming field ($\ket{gg}\leftrightarrow\ket{+}$)
or indirectly through the bright state 
($\ket{gg}\leftrightarrow\ket{-}\leftrightarrow\ket{+}$), to which it is 
coupled via the exchange term (Fig.~\ref{FigSchematic}b). 

Our measurement setup allows us to drive the system from 
either the forward or reverse direction 
($\alpha$ and $\beta$ driving, respectively, in Fig.~\ref{FigSchematic}a), 
with the reflected and transmitted fields 
simultaneously detected at both sides (refer to the Supplementary Material
for extra details on the measurement setup). When driving in the forward direction, 
both channels to populate the quasi-dark state $\ket{+}$ interfere 
constructively, giving rise to an excitation of $\ket{+}$. 
Neglecting non-radiative decay and dephasing ($\gamma_{nr}=\gamma_{\phi}=0$), the resulting steady state 
solution for the  density operator of the qubits can be found 
analytically~\cite{Muller2017a} as 
$\rho_{st}=(1/3)|gg\rangle\langle gg|+(2/3)|+\rangle \langle +| + O(\delta^2)$ 
for intermediate driving powers $\delta^2\overbar{\gamma_r} \ll p \ll 2\overbar{\gamma_r}$. 
Under these conditions, the system is predominantly trapped in the 
quasi-dark state $\ket{+}$ and is therefore partially transparent to the incident signal,
due to the extremely low saturability of  $\ket{+}$.

If the system is driven in the reverse direction, both channels 
interfere destructively, the quasi-dark state remains unpopulated
and the steady state solution is given by 
$\rho_{ss}=\ket{gg}\bra{gg} + O(\delta^2)$ for powers $p\ll 2\overbar{\gamma_r}$. 
In this case, the incoming signal is reflected by the bright state and the two-qubit 
system behaves as a mirror.

\begin{figure}[t] \centering
  \includegraphics[width=\columnwidth]{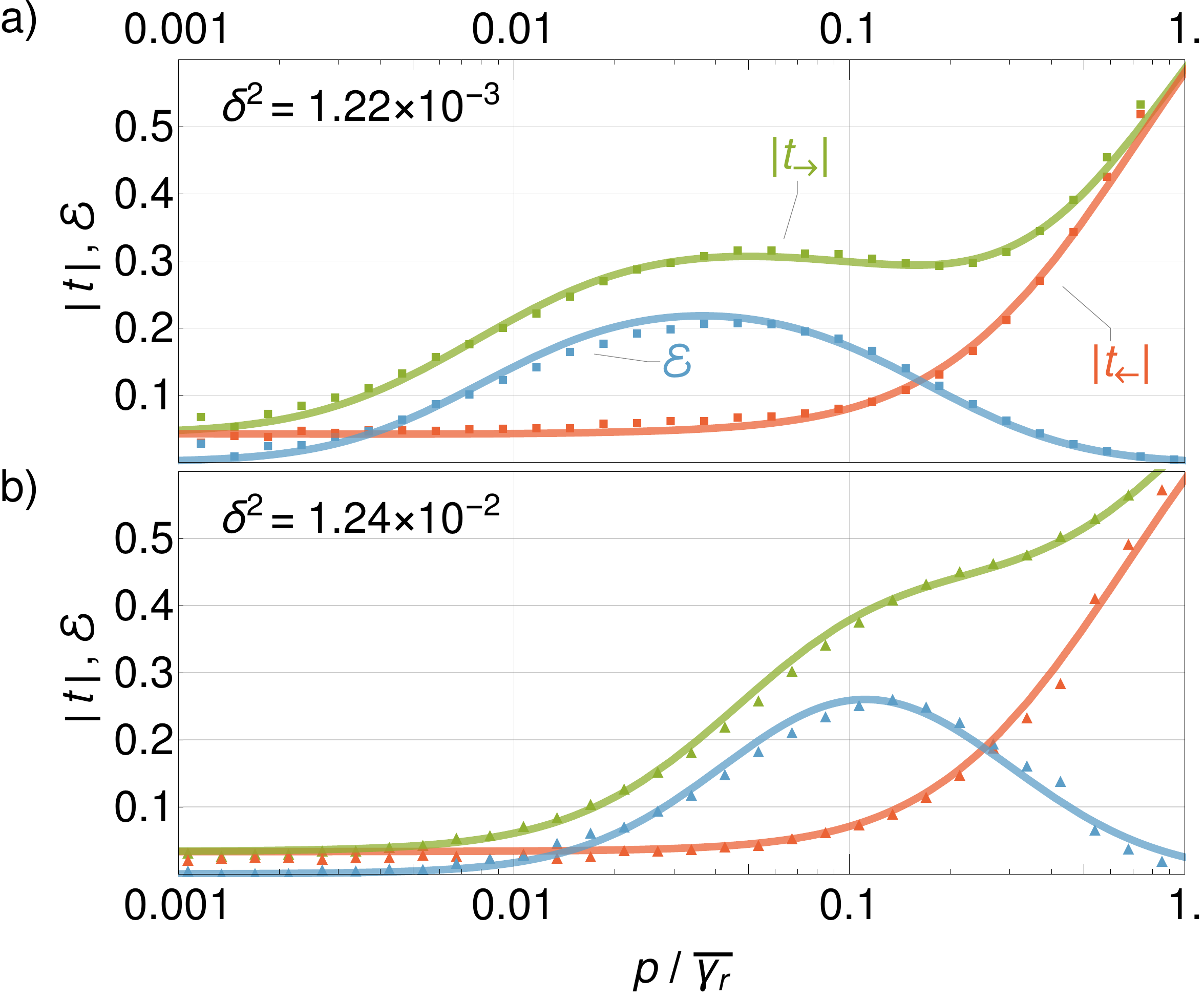}
   \caption{Nonreciprocity dependence on power. Experimental data 
   (points) and theoretical fits (solid lines) for the forward 
   driving transmission amplitude $|t_\rightarrow|$ (green), 
   reverse driving transmission amplitude $|t_\leftarrow|$ (red) and diode efficiency $\mathcal{E}$ (blue). 
   $\mathcal{E} \equiv |t_\rightarrow|(|t_\rightarrow|-|t_\leftarrow|)/(|t_\rightarrow|+|t_\leftarrow|)$
   measures the nonreciprocal behavior of the system as well as 
   its forward transmitting capabilities.
   The system is tuned to its optimal nonreciprocal configuration for 
   two different detunings:  a) $\delta^2 \simeq 0.001$, b) 
   $\delta^2 \simeq 0.01$. In both cases
   the device behaves reciprocally and reflects most of the incoming radiation
   at the low power regime $p/\overbar{\gamma_r}\ll\delta^2$.
   However, as the power increases past the onset of the diode 
   regime, indicated by $\delta^2$, saturation of the quasi-dark state 
   allows for an increase in $|t_\rightarrow|$, while $|t_\leftarrow|$
   remains unchanged. 
   } 
	\label{FigPowerSweep}
\end{figure}

To illustrate the mechanism of nonreciprocal transmission, 
we  tune the system to its optimal nonreciprocal configuration (up 
to experimental uncertainties) for two different values of the parameter 
$\delta$: $\delta^2 \simeq 10^{-2}$ and $\delta^2 \simeq 10^{-3}$, 
corresponding to driving frequencies $\omega_{d}\simeq 8.8358\textrm{ GHz}$ 
and $\omega_{d}\simeq 8.6188\textrm{ GHz}$, respectively. 

By controlling the driving power, we are able to probe three 
characteristic regimes of the device, featured in Fig.~\ref{FigPowerSweep}:
\begin{itemize}
	\item In the low power regime, $p/\overbar{\gamma_r}\ll\delta^2$, 
	the device behaves reciprocally, reflecting most of the incoming radiation. 
	In this regime, the degree of transmission suppression is only limited by the 
	qubits' decoherence and relaxation rates, and by the accuracy of qubit tuning
	to ensure that $\omega_2 = \omega_d$.
	\item In the intermediate power regime,   
	$\delta^2\ll p/\overbar{\gamma_r}\ll 1$, the transmission 
	amplitude in the forward direction $t_\rightarrow$ increases and 
	features the characteristic plateau predicted by theory~\cite{Dai2015,Muller2017a}. 
	The transmission amplitude in the reverse direction $t_\leftarrow$ 
	remains near zero independently of the value of $p/\overbar{\gamma_r}$ and the system behaves nonreciprocally. 
	\item In the high power regime, $p/\overbar{\gamma_r}\gg 1$, 
	the bright state saturates, regardless of the driving direction, 
	and the system returns to its reciprocal behavior.
\end{itemize}

\begin{figure}[t]
	\centering
	\includegraphics[width=\columnwidth]{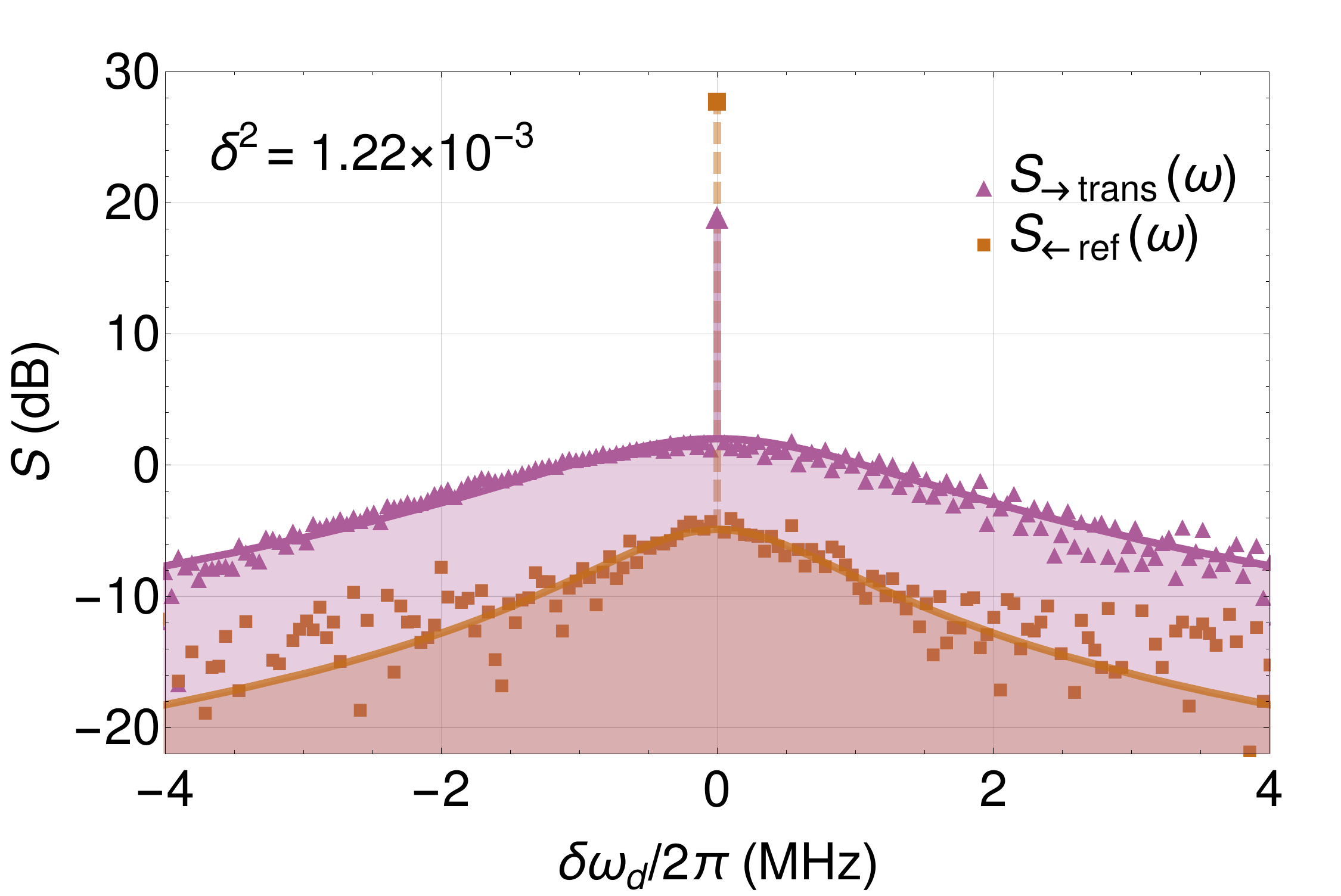}
	\caption{Power spectral densities of the forward driving transmitted 
		field ($S_{\rightarrow \textrm{trans}}(\omega)$) and the reverse driving reflected 
		field ($S_{\leftarrow \textrm{ref}}(\omega)$). Both spectra were taken 
		after tuning the device to its optimum nonreciprocal configuration for
		$\delta^2 \simeq 0.001$ (as in Fig.~\ref{FigPowerSweep}a) and setting the driving power
		such that the diode efficiency $\mathcal{E}$ is maximum
		(corresponding to $p/\overbar{\gamma_r}\simeq 0.05 $ in Fig.~\ref{FigPowerSweep}a).  
		Both scattered fields travel through the same amplification chain.
		The $\delta$-like peaks at the driving frequency $\omega_d$
		correspond to the elastically scattered fields. 
		Radiation inelastically scattered off the quasi-dark state $\ket{+}$
		generates an additional broader peak, which we fit to Lorentzians
		of width $\gamma_{\rightarrow \textrm{trans}} = 2.92\textrm{ MHz}$ and 
		$\gamma_{\leftarrow \textrm{ref}} = 1.76\textrm{ MHz}$ (solid lines).
		When driving the system in the forward direction, part of the population
		gets trapped in the quasi-dark state, consistent with a greater
		inelastically scattered radiation power: 
		$ \int\!S_{\rightarrow \textrm{trans}} \, d\omega \gg \int\!S_{\leftarrow \textrm{ref}} \, d\omega$. } 
	\label{FigPSD88}
\end{figure}

In order to provide a metric of the isolation capabilities of the quantum diode, 
we calculate the \emph{diode efficiency} 
$\mathcal{E}\equiv|t_\rightarrow|(|t_\rightarrow|-|t_\leftarrow|)/(|t_\rightarrow|+|t_\leftarrow|)$ 
used in Ref.~\cite{Dai2015} which, in the ideal case of identical 
qubits and no decoherence, coincides with the definition of efficiency 
used in Refs.~\cite{Fratini2014,Dai2015, Fratini2016}. 
In spite of relatively low dephasing and non-radiative decay rates 
($\gamma_{\phi}, \gamma_{nr} < 0.5\%\, \gamma_{r}$ for both qubits), the maximum diode efficiency appears 
to be limited to $\simeq 0.27$, well below its ideal value~\cite{Muller2017a} of $2/3$
(see Fig.~\ref{FigPowerSweep}).
This illustrates an experimental challenge in the realization of the quantum diode:
since the nonreciprocal behavior relies on populating
the quasi-dark state $\ket{+}$, the transition rate
relevant for the system dynamics is
$\Gamma_+=\delta^2\overbar{\gamma_r}\ll\overbar{\gamma_r}$.
In our experiment, the dephasing and dissipation 
rates, $\gamma_\phi$, $\gamma_{nr}$, are of the same order of magnitude of $\Gamma_+$. This renders the effect of decoherence much more 
significative compared to single-qubit phenomena,
whose dynamics evolve at the much faster rate $\gamma_r$.

\begin{figure}[t!]
	\centering
	\includegraphics[width=\columnwidth]{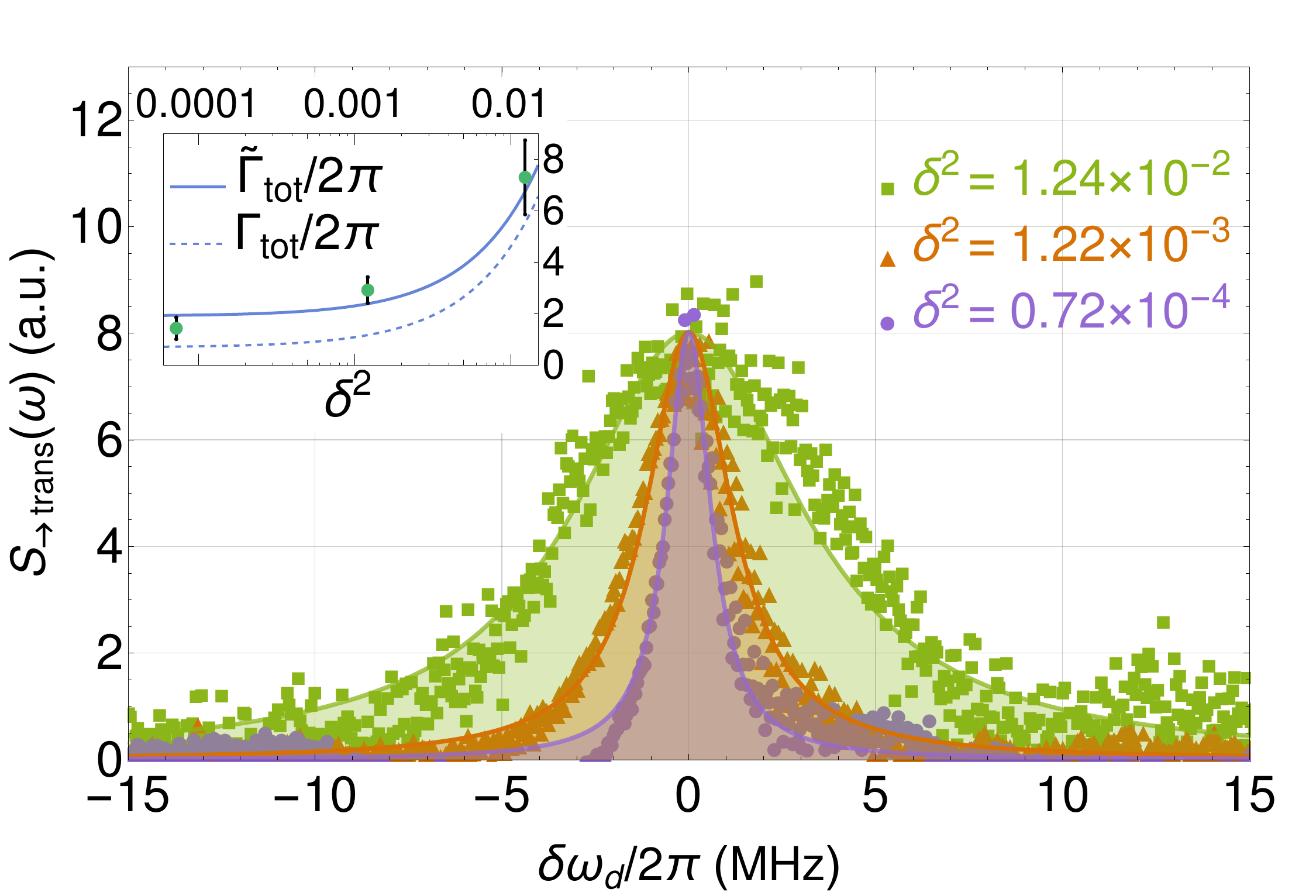}
	\caption{ Power spectral densities of the inelastically scattered transmitted field.
		The system is driven in the forward direction for different detuning regimes 
		($\delta^2 \simeq 0.01, 0.001, 0.0001$).
		In every case, the spectra were taken at the 
		optimum atomic detunings and driving powers that
		maximize the diode efficiency $\mathcal{E}$. 
		The solid lines are fits to Lorentzians 
		of width 1.43~MHz, 2.92~MHz and 7.30~MHz, in increasing order of $\delta^2$.
		As predicted by theory, the linewidth 
		$\Gamma_{\textrm{FWHM}}\equiv 2(3\delta^2\overbar{\gamma_{r}}+2\gamma_\phi+\gamma_{nr})$
		of the scattered field increases linearly with $\delta^2$ (see inset).  
		However, in the experiment the width of the inelastically 
		scattered radiation is broader than expected: 
		$\tilde\Gamma_\textrm{tot}=\Gamma_\textrm{tot}+\Gamma_\textrm{exc}$,
		with $\Gamma_\textrm{exc}/2\pi=1.22\textrm{ MHz}$.          
		The elastic part of the scattered radiation at $\delta\omega_d = 0\textrm{ MHz}$ 
		has been omitted and the peaks are scaled for clarity.}       
	\label{FigPSD88and86}
\end{figure}

Further insights into the role of the quasi-dark state 
can be obtained by measuring the full spectrum of the elastically and 
inelastically scattered radiation. Notably, the measurement 
of the power spectral densities, in addition to a $\delta$-like peak due 
to elastically (Rayleigh) scattered radiation (Fig.~\ref{FigPSD88}), 
features an additional broader peak, which we identify with radiation 
inelastically scattered off the quasi-dark $\ket{+}$ state. The measured power spectrum agrees with 
our expectation of the total scattered power being a measure of the 
population of the quasi-dark state: as clearly seen from the measurement (Fig.~\ref{FigPSD88}), the scattered 
power is much greater when the system is driven in the forward direction.

Being able to control $\delta$, we can tune the linewidth 
of the dark state emission, $\Gamma_+=\delta^2\overbar{\gamma_r}$. 
Fig.~\ref{FigPSD88and86} shows the power spectral densities 
of scattered radiation for three values of $\delta$ (here the elastic part has been omitted for clarity). 
As expected, the linewidth of the fluorescence spectra increases linearly with $\delta^2$,
 following the increase in the decay rate of the dark state.

Our theoretical estimates predict that the width of the emission peak 
results from a combination of the dark state linewidth $\Gamma_+$
and broadening due to non-radiative and dephasing contributions. 
In the optimal diode conditions, 
the linewidth of the transmitted field when driving the system
in the forward direction can be found analytically as 
$\Gamma_\textrm{FWHM} = 2(3\Gamma_+ + 2 \gamma_{\phi} + \gamma_{nr})$
(see Supplementary Material). 
In the experiment, the width of the inelastically scattered radiation 
is wider than predicted by $1.22$~MHz. 
This indicates an additional source of noise  (presumably of technical origin) 
which could be mixed with the detected signal and results in an additional broadening of the scattered field.

By incorporating quantum-limited Josephson parametric amplifiers  
into our detection lines \cite{Eichler2014}, we can measure time-domain single-shot data 
of the scattered fields and calculate its statistics. 
Our results show that the in-phase noise
is higher when driving the system in the forward direction
than in the reverse direction (see Supplementary Material). 
This is consistent with the statistics produced by replacing the system 
with a simple stochastic mirror, as theoretically predicted in Ref.~\cite{Muller2017a}.

In conclusion, we experimentally realized a passive quantum nonreciprocal device 
comprised of a minimal number of constituents. At least two localized quantum emitters are required to break 
structural symmetry in 1D space, while a two-level atom is the
simplest system presenting a nonlinear quantum behavior. 
The nonreciprocity relies on the interplay of the exchange 
interaction and the collective decay of quantum emitters leading to population 
trapping into an entangled quasi-dark state for a preferred driving direction. It is instructive to note that our device breaks some of the fundamental bounds derived for classical nonlinear devices~\cite{Sounas2017,Sounas2018} but is not immune to the dynamic reciprocity limitations~\cite{Shi2015}.
While not yet sufficient for practical applications, our results open a path for 
the realization of more efficient nonreciprocal devices with multiple coherent qubits. The demonstrated  mechanism of population trapping is 
also valuable for the development of protocols of remote entanglement stabilization.

\acknowledgements{ 
We thank Alexandre Roulet for useful discussions of our results and Shanhui Fan for helpful advise on dynamical reciprocity. We also thank  Andrea Al\`u for reading our manuscript and providing valuable insights into nonlinear nonreciprocity. We acknowledge Andreas Wallraff and the ETHZ  Qudev  team  for  providing us with the  parametric  amplifier. 
This work was supported by the Australian Research Council under the Discovery and
Centre of Excellence funding schemes (project numbers DP150101033, DE160100356, and CE110001013), and UQ Foundation Research Excellence Award. M.W. acknowledges support from the European Research Council (ERC) under the Grant Agreement 648011. 
}

\bibliography{biblio}

\clearpage

\appendix

\section{Measurement setup}

\begin{figure}[b]
	\centering
  \includegraphics[width=\columnwidth]{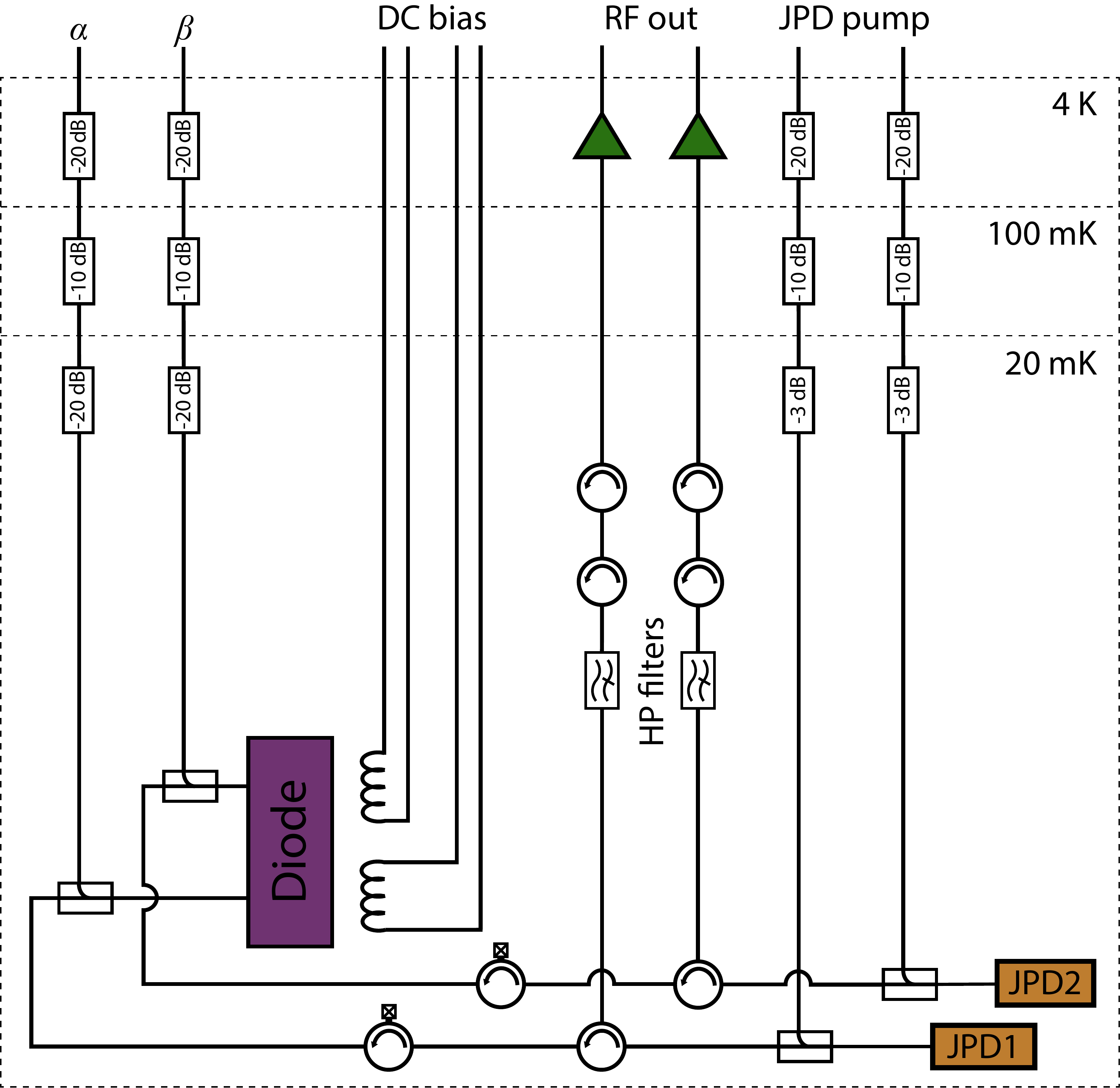}
   \caption{Cryogenic setup. Two qubits embedded in a 1D waveguide (purple box),
   which can be driven from either of the input lines $\alpha$, $\beta$. 
   After amplification at the Josephson parametric dimers  (orange boxes)
   and high-electron-mobility transistors (green triangles),
   the transmitted and reflected fields can be simultaneously detected at RF out.
    }
	\label{FigCryoSetup}
\end{figure}

The nonreciprocal device, consisting of two superconducting qubits embedded 
in a 1D waveguide, is mounted on the 20 mK stage of a dilution refrigerator (purple
box in Fig.~\ref{FigCryoSetup}). Two superconducting coils mounted 
around the waveguide serve to control the Josephson inductance of each
of the qubits, which in turn determine their transition
frequencies $f_{ge}$.

The device can be driven 
from any of the two input ports $\alpha$, $\beta$, which
are interfaced with the waveguide through the coupled ports
of two directional couplers. The reflected
and transmitted signals are then collected via the through
port of the directional couplers, which allows us to detect them 
simultaneously. The two output signals are then parametrically amplified via
two Josephson parametric dimers (JPD) \cite{Eichler2014}
(orange boxes in Fig.~\ref{FigCryoSetup})
and further amplified via two high-electron-mobility transistor (HEMT)
amplifiers. 

\section{Single qubit SLH description}

For a single atom in a waveguide, coupled symmetrically to 
both propagation directions, and driven with a coherent field 
from both sides, we can write, in analogy to Ref.~\cite{Muller2017a},
\begin{align}
	H_\textrm{tot} &= -\frac12 \delta\omega \sigma_{z} + \frac{1}{2\ii} \sqrt{\frac{\gamma_{r}}{2}} \bigl( ( \alpha + \beta ) \sigma_{+}  - \text{h.c.} \bigr) \,,\label{eq:H1}\\
	a_{\text{out}} &= \sqrt{\frac{\gamma_{r}}{2}} \sigma_{-} + \alpha \quad \,, \quad	b_{\text{out}} = \sqrt{\frac{\gamma_{r}}{2}} \sigma_{-} + \beta \,,
\end{align}
where $\gamma_{r}$ is the total radiative decay rate into the waveguide modes, $\alpha$ is the strength of the coherent drive from the left, and $\beta$ is the driving field amplitude from the right. 
Here, Eq.~\eqref{eq:H1} is written in a frame  rotating at the drive 
frequency $\omega_{d}$, and $\delta\omega = \omega_{q}- \omega_{d}$ 
is the detuning between the atomic transition frequency $\omega_{q}$ and the drive.
The master equation, including non-radiative decay and dephasing on the atom, is then
\begin{align}
	\dot \rho =& -\ii \comm{H_\textrm{tot}}{\rho} + \diss{a_{\text{out}}}\rho + \diss{b_{\text{out}}}\rho \nn\\
		&+ \gamma_{nr} \diss{\sigma_{-}}\rho + \frac12 \gamma_{\phi} \diss{\sigma_{z}}\rho \,,
\end{align}
with the non-radiative decay rate $\gamma_{nr}$ and the dephasing rate $\gamma_{\phi}$.

To calculate the transmittance, we first find the steady-state of the master equation, $\dot {\bar \rho} =0$.
The transmission amplitude when driving from the left is then
\begin{align}
	t &= \Tr\l\{ a_{\text{out}} \bar\rho\r\} / \alpha \nn\\
		&= 1-\frac{\gamma_{r}}{2\gamma_{2}} \frac{1-\ii {\delta\omega}/{\gamma_{2}}}{1+({\delta\omega}/{\gamma_{2}})^{2} + { 2\abss\alpha\gamma_{r}}/{\gamma_{1} \gamma_{2}}} \,, \label{eq:t1}
\end{align}
where we defined the total atomic decay rate $\gamma_{1} = \gamma_{r} + \gamma_{nr}$ and dephasing rate $\gamma_{2} = \frac12 \gamma_{1} + \gamma_{\phi}$, 
and $|\alpha|^2$ is the power of the incoming drive.

\section{Single qubit spectroscopy}
\label{singleQubitSpec}

We characterize the radiative $\gamma_r$,
non-radiative $\gamma_{nr}$, and dephasing $\gamma_\phi$ rates
of each individual qubit by performing spectroscopic measurements. 
In order to achieve this, we detune one of the qubits below the cutoff 
of the waveguide and set the target qubit to frequency $\omega_q$.
We then sweep a low power microwave tone $\omega_d$ around $\omega_q$
and detect the transmitted electric field $E_t$. Next, we
calculate the transmission amplitude $t=E_t/E_i$, 
where $E_i$ is the amplitude of the incident electric field. 
The transmittance is then least-squares fitted to Eq.~\eqref{eq:t1} to extract 
all relevant parameters for each qubit individually. 
We repeat this procedure for a range of target frequencies for 
each qubit in order characterize their operational points.
The radiative decay rates $\gamma_r$ are found to vary between $2\pi\times 60$~MHz and $2\pi\times 85$~MHz
for qubit frequencies between 8.5~GHz and 9.5~GHz, 
whereas $\gamma_{\phi}, \gamma_{nr} < 0.5\% \gamma_{r}$
throughout this range. See Table~\ref{TableSingleQubitSweep} for some particular values.

\begin{table}[]
\centering
\begin{tabular}{ccc}
     & \quad  8.6 GHz  \quad & \quad 8.8 GHz \quad \\ \hline
$\gamma_{r,1}/2\pi$ (MHz)  &   71.3039   &  62.4261      \\ 
$\gamma_{r,2}/2\pi$ (MHz)  &   72.4299   &  73.1158      \\ 
$\gamma_{\phi}/2\pi$ (kHz)  &   211.4   &  74.7      \\ 
$\gamma_{nr}/2\pi$ (kHz)  &   191.1   &  64.0      \\ 
\hline
\end{tabular}
\caption{Least-squares fit results for qubits 1 and 2 
tuned to 8.6 GHz and 8.8 GHz. 
We use the methods outlined in App.~\ref{singleQubitSpec}
to fit $\gamma_{r,i}$ and $\gamma_{nr}+2\gamma_{\phi}$. 
We assume equal dephasing and non-radiative decay rates for both qubits. 
We then use the two-qubit transmission amplitude expressions 
found in App.~\ref{twoQubit} to get individual values
for $\gamma_{nr}$ and $\gamma_{\phi}$ when the 
device is configured in the diode regime.
}
\label{TableSingleQubitSweep}
\end{table}

\section{Two-qubit master equation}
\label{twoQubit}

Following Ref.~\cite{Muller2017a}, we write the master equation for the 
two-qubit density matrix 
\begin{align}
	\dot{\rho} = \mathcal{L}\rho\equiv  & -\ii \comm{H_T}{\rho} + \diss{a_{\text{out}}}\rho + \diss{b_{\text{out}}}\rho \nn \\ 
	&  +\gamma_{nr}\l(\diss{\sigma_-^{(1)}}\rho+\diss{\sigma_-^{(2)}}\rho\r) \nn\\
	& + \gamma_\phi \l(\diss{\sigma_z^{(1)}}\rho+\diss{\sigma_z^{(2)}}\rho\r),
	\label{eqME}
\end{align}
where $\diss{X}\rho=X\rho X\hc-\tfrac12(X\hc X\rho +\rho X\hc X)$, and 
\begin{align*}
	H_T =& \,H_1 + H_2 -\ii 
	\big( \alpha L_1\hc - \alpha\cc L_1 \big)/2 -\ii 
	 \big( \beta L_2\hc - \beta\cc L_2 \big)/2 \nn\\
		&-\ii 
		 \big( \ee^{\ii \phi} L_2\hc (L_1 + \alpha) - \ee^{-\ii \phi} (L_1\hc + \alpha\cc) L_2 \big)/{2} \nn\\
		&-\ii 
		\big( \ee^{\ii \phi} L_1\hc (L_2 + \beta) - \ee^{-\ii \phi} (L_2\hc + \beta\cc) L_1 \big)/{2} \,, \\
	a_\textrm{out}=& \,(\alpha+L_1)\ee^{\ii \phi} + L_2  \,,\\
	b_\textrm{out}=& \,(\beta+L_2)\ee^{\ii \phi} + L_1  \,.
\end{align*}
Here $\alpha,\beta$ are the amplitudes of the right-
and left-moving fields, respectively, and the operators 
$a_{\text{out}}$ and $b_{\text{out}}$ represent the right- and left-moving output fields.  
The input field is at frequency $\omega_d$, $\phi$ is the phase shift acquired by 
the drive when traveling between the atoms,
and $L_i$, $H_i$ form part of the SLH triplet \cite{CombKercSaro16} described by
\begin{align*}
	H_k =  - \omega_k\, \sigma_z^{(k)} /2\,,\quad   
	L_k = \sqrt{{\gamma_{r,k}}/2}\, \sigma_-^{(k)},
\end{align*}
where $k=1,2$ indexes the atoms, $\omega_k$ is the eigenfrequency of 
atom $k$ in the frame rotating at $\omega_d$, and $\gamma_{r,k}$ is its radiative decay rate. 
We defined the atomic lowering operator as $ \sigma_-={\ket{g}}{\bra{e}}$.  

The right-moving steady-state output field is then found as
$\langle a_{\text{out}} \rangle^{(\alpha,\beta)}_{SS} =  \text{Tr}\l\{  a_\textrm{out} \rho^{(\alpha,\beta)}_{SS} \r\}$,
where $\rho^{(\alpha,\beta)}_{SS}$ is the steady state solution of Eq.~\ref{eqME}, $\dot\rho_{SS}^{(\alpha,\beta)}=0$, with input amplitudes $\alpha,\beta$.
Finally, we calculate the forward-driven transmission amplitude as 
$t_f = \langle a_{\text{out}} \rangle^{(\alpha,0)}_{SS}/\alpha$.
Similarly, the reverse-driven transmission amplitude equates 
to $t_r = \langle b_{\text{out}} \rangle^{(0,\beta)}_{SS}/\beta$.

\section{Bright and dark states decay rates for $\gamma_{r,1}\neq\gamma_{r,2}$}

Following the notation of Ref.~\cite{Muller2017a}, we write the dissipative parts of the
SLH dissipators as
\begin{align*}
	\bar L_{1} &= \frac12 \l\{ \sqrt{\gamma_{r,1}} \l( \sigma_{-}^{D} - \sigma_{-}^{B} \r) + \sqrt{\gamma_{r,2}} \ee^{\ii \phi} \l( \sigma_{-}^{D} + \sigma_{-}^{B} \r) \r\} \,,\nn\\
	\bar L_{2} &= \frac12 \l\{ \sqrt{\gamma_{r,2}} \l( \sigma_{-}^{D} + \sigma_{-}^{B} \r) + \sqrt{\gamma_{r,1}} \ee^{\ii \phi} \l( \sigma_{-}^{D} - \sigma_{-}^{B} \r) \r\} \,, 
\end{align*}
where we assumed the general case of $\gamma_{r,1} \neq \gamma_{r,2}$, 
and we used the dark and bright state annihilation operators
\begin{align*}
	\sigma_{-}^{D} &= \frac{1}{\sqrt2} \l( \sigma_{-}^{(2)} + \sigma_{-}^{(1)} \r), \\
	\sigma_{-}^{B} &= \frac{1}{\sqrt2} \l( \sigma_{-}^{(2)} - \sigma_{-}^{(1)} \r).
\end{align*}
In the experimentally relevant limit where $\phi=\pi - \delta$, $\delta \ll \gamma_{r,k}$, and $\gamma_{r,1}\approx \gamma_{r,2}$,
we can rewrite the the sum of dissipators in the master equation as
\begin{align}
	\diss{\bar L_{1}}\rho + \diss{\bar L_{2}}\rho \simeq \gamma_{D}\diss{\sigma_{-}^{D}}\rho + \gamma_{B}\diss{\sigma_{-}^{B}}\rho \,,
\end{align}
with $\gamma_D = \delta^2 \overbar{\gamma_r}/2$ and $\gamma_B = 2 \overbar{\gamma_r}$,
and where we defined $\overbar{\gamma_r} \equiv \sqrt{\gamma_{r,1} \gamma_{r,2}}$.

\section{Linewidth of the transmitted field}

The linewidth of the dark state emission is given by the lifetime of the 
dark state population $p_{D}$, i.e., the rate at which it decays towards its equilibrium state. 
In the diode regime, the dark state is inverted, in that $p_{D}>1/2$. Without nonradiative decay or additional dephasing, from Ref.~\cite{Muller2017a} we find
the decay rate of the dark state into the ground-state in the diode regime as 
$\gamma_{D} = \delta^{2} \overbar{\gamma_{r}} /2$. 
In the same regime, the excitation rate of exciting the 
ground state into the dark state is $\gamma_{\uparrow D} = 2\gamma_{D}$, 
leading to the ideal steady-state dark state population of 
$p_{D} = \gamma_{\uparrow D} / (\gamma_{\uparrow D} + \gamma_{D})  = 2/3$. 

In the non-ideal case, assuming symmetric nonradiative and dephasing as in Eq.~\eqref{eqME}, 
and focusing on their contributions to the dark state decay rate, 
one finds 
\begin{align}
	\gamma_{nr} \l( \diss{\sigma_{-}^{(1)}}\rho + \diss{\sigma_{-}^{(2)}}\rho \r) &= \gamma_{nr} \l( \diss{\sigma_{-}^{D}}\rho +\diss{\sigma_{-}^{B}}\rho \r) \,, \\
	\gamma_{\phi} \l( \diss{\sigma_{z}^{(1)}}\rho + \diss{\sigma_{z}^{(2)}}\rho \r) 
		&= \frac{\gamma_{\phi}}{2} \diss{\sigma_{z}^{1}+\sigma_{z}^{2}}\rho +2 \gamma_{\phi} \diss{\sigma_{-}^{D}}\rho \,, 
\end{align}
leading to the total decay rate of the dark state $\gamma_{D}' = \gamma_{D} +\gamma_{nr} + 2\gamma_{\phi}$. The dark state lifetime is then
$\Gamma_{D} = \gamma_{D}' + \gamma_{\uparrow,D} = 3\gamma_{D} + \gamma_{nr} + 2\gamma_{\phi}$ and the experimentally measured total linewidth of the emitted radiation is given by 
$\Gamma_\textrm{FWHM} = 2\Gamma_{D}$.

\section{Output field statistics in the diode configuration}

\begin{figure}[t]
	\centering
  \includegraphics[width=\columnwidth]{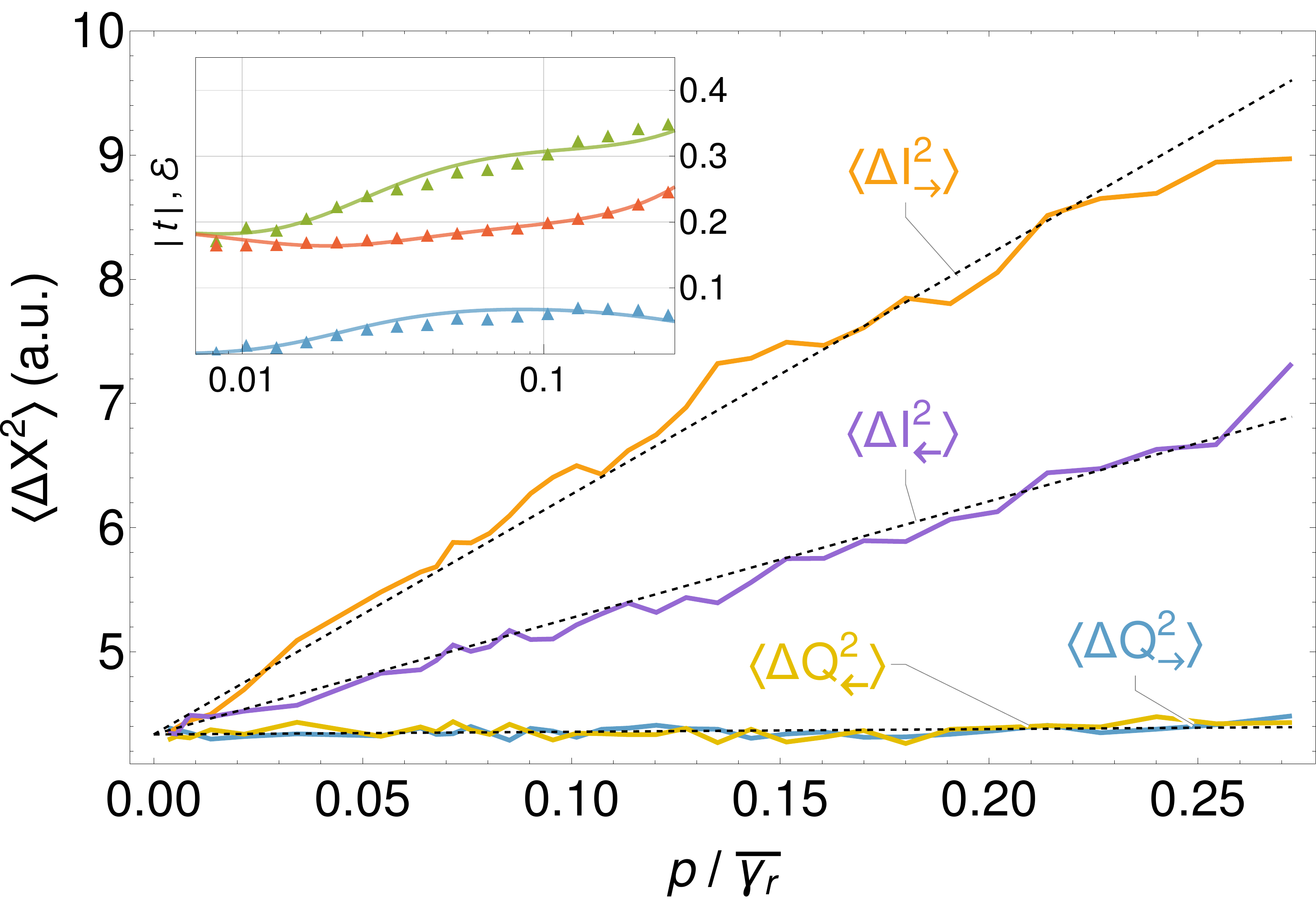}
  \caption{ Variance of the single-shot detected output fields with increasing power. 
  The device was prepared at the optimum diode configuration for $\delta^2 \simeq 0.01$.
  At each power, $2^{18}$ single shots of the forward and reverse driving 
  heterodyne IQ signal were detected at a 100 MHz rate. The statistics of the collected
  data features a linear dependence with power in the in-phase noise $\mean{\Delta I_\rightarrow^2}$,
  consistent with a constant population of the quasi-dark state. The quadrature noise
  remains at a constant value, regardless of the driving direction,
  $\mean{\Delta Q^2_\rightarrow} = \mean{\Delta Q^2_\leftarrow} = \sigma_w^2 = \sigma_\textrm{qn}^2 + \sigma_{\textrm{tech}}^2$,
  where $\sigma_{\textrm{tech}}^2$ and $\sigma_\textrm{qn}^2$ are the technical and quantum noise
  contributions to the signal, respectively.
  Due to a limitation in the frequency tuning range of the Josephson parametric amplifiers, the measurement was performed in 
  a separate cooldown for a larger distance between qubits $d=44.95\textrm{ mm}$ and 
  at a lower driving frequency $\omega_d/2\pi\simeq 7.398\textrm{ GHz}$.
  At this regime, the diode efficiency $\mathcal{E}$ was found to be limited to $0.07$. 
  Inset: Experimental data (points) and theoretical fits (solid lines) for the forward (green)
  and reverse (red) transmission amplitudes. The diode efficiency $\mathcal{E}$ is plotted in blue.}    
	\label{FigNoise}
\end{figure}

When driving the diode in the forward direction and powers $p/\overbar{\gamma_r} \ll 1$,
the system is confined to the $\{\ket{G}, \ket{D}\}$ manifold (up to order $\delta^2$). 
The two-atom system can be found in the ground, reflecting state with probability $(1-P_D)$
or in the quasi-dark, transparent state with probability $P_D$.
As demonstrated in Ref.~\cite{Muller2017a}, the statistics of the 
scattered fields in the diode regime can be replicated 
by a \emph{flapping mirror} model, i.e., by a system composed 
of a stochastic mirror that can flip into the path of the 
signal (reflecting state, $X=0$) with probability $(1-P_D)$,
or out of the path (transmitting state, $X=1$) with probability $P_D$. 
In this scenario, the transmitted in-phase heterodyne signal $I$ can
be modeled as $I(t) = S(t) + w(t)$,
where $S(t) = \alpha X(t)$ is the transmitted signal, which
depends on the driving amplitude $\alpha$ and the state of the mirror $X$,
and $w(t)$ encompasses both quantum and technical noise contributions.
Assuming uncorrelated white noise, such that $\bar{w}=0$, 
the variance of the in-phase signal is given by 
\begin{align*}
\mean{\Delta I^2} &= \mean{(S(t) + w(t) -\bar{S})^2}  \\
                  &= \mean{(S(t)-\bar{S})^2} + \mean{w(t)^2} \\
                  &= \sigma_S^2 + \sigma_w^2.  
\end{align*}
Since $S(t)=\alpha X(t)$, $\sigma_S^2 = |\alpha|^2 \sigma_X^2 = |\alpha|^2 P_D (1-P_D)$,
which gives
\begin{align*}
\mean{\Delta I^2} &= |\alpha|^2 P_D (1-P_D) + \sigma_w^2.  
\end{align*}
On the other hand, the quadrature signal remains unchanged regardless
of the state of the mirror $Q(t) = w(t)$,
and hence $\mean{\Delta Q^2} = \sigma_w^2$. In both the forward- and reverse-driving configurations, 
the noise in the reflected and transmitted signals is the same since $R(t) = 1 - S(t)$.


We incorporate quantum-limited Josephson parametric amplifiers  
into our detection lines in order to measure time-domain single-shot data 
of the scattered fields and compare our results to the
flapping mirror model (see Fig.~\ref{FigNoise}). 
When forward-driving the system with powers $p/\overbar{\gamma_r} \ll 1 $, 
the quasi-dark state population $P_{D,\rightarrow}$ reaches a constant non-zero value (up to order 
$\delta^2$), and hence we observe that
the in-phase noise scales linearly with power: $\mean{\Delta I^2_\rightarrow} = C_1 + C_2 \, p$. 
As expected, the quadrature
noise remains at its constant value for both driving directions
$\mean{\Delta Q^2_\rightarrow} = \mean{\Delta Q^2_\leftarrow} = \sigma_w^2 = \sigma_\textrm{qn}^2 + \sigma_{\textrm{tech}}^2$,
where $\sigma_{\textrm{tech}}^2$ and $\sigma_\textrm{qn}^2$ are the technical and quantum noise
  contributions to the signal, respectively.
The in-phase noise for the reverse driving scattered field $\mean{\Delta I^2_\leftarrow}$ also increases
linearly with power, although at a different rate, 
consistent with a smaller population $P_{D,\leftarrow} < P_{D,\rightarrow}$
of the quasi-dark state and with our simulations.

\end{document}